\documentstyle[sprocl]{article}

\def\Journal#1#2#3#4{{#1} {\bf #2}, #3 (#4)}

\def\NPB{{\em Nucl. Phys.} B}
\def\PLB{{\em Phys. Lett.}  B}
\def\PRL{\em Phys. Rev. Lett.}

\def\be{\begin{equation}}
\def\ee{\end{equation}}
\def\bea{\begin{eqnarray}}
\def\eea{\end{eqnarray}}
\begin{document}

 \begin{flushright}
  \begin{tabular}{l}
  NORDITA--96--30--P\\
  SLAC-PUB-7176\\
  hep-ph/9605375\\
  May 1996
  \end{tabular}
  \end{flushright}
  \vskip0.5cm

\title{RESUMMATION OF THRESHOLD CORRECTIONS IN QCD TO POWER ACCURACY:
THE DRELL-YAN CROSS SECTION AS A CASE STUDY\footnote{
Research supported 
by the Department of Energy under contract DE-AC03-76SF00515.}
}

\author{ M. BENEKE
        }

\address{Stanford Linear Accelerator Center,\\ 
Stanford University, Stanford, CA94309, U.S.A.}

\author{ V.M. BRAUN }

\address{NORDITA, Blegdamsvej 17, DK-2100 Copenhagen, Denmark}


\maketitle\abstracts{
Resummation of large infrared logarithms in perturbation theory 
can, in certain circumstances, enhance the sensitivity to 
small gluon momenta and introduce spurious nonperturbative
contributions. In particular, different procedures -- equivalent 
in perturbation theory -- to organize this resummation can 
differ by $1/Q$ power corrections. The question arises whether one 
can formulate resummation procedures that are explicitly 
consistent with the infrared behaviour of finite-order Feynman diagrams.
We explain how this problem can be treated and resolved
in Drell-Yan (lepton pair) production and briefly discuss more 
complicated cases, such as top quark production and 
event shape variables in the $e^+e^-$ annihilation.}

\vspace{1cm}

\begin{center}
{\em To appear in the Proceedings of \\
         Second Workshop on Continuous Advances in QCD\\
         (Minneapolis, U.S.A., March 1996)\\
          and\\
          1996 Zeuthen Workshop on Elementary Particle Theory:\\         
         QED and QCD in Higher Orders\\
         (Rheinsberg, Germany, April 1996)}
\end{center}
\newpage

\section{Introduction}

The potential to study ``new physics'' at the next generation of accelerators
will to a large extent depend on the ability to control the strong 
interaction background. Hence the present interest in making
QCD predictions for hard processes as quantitative as possible,
with an increasing understanding of the importance to study higher-twist
effects which are suppressed by powers of the large momentum.
In general, higher-twist effects reflect the ``leakage''
of contributions from large distances into the process of interest.
The theoretical status of these corrections is well-established in 
the total $e^+ e^-$ annihilation cross section and in deep inelastic
scattering (DIS) (and in related quantities), 
where dispersion relations relate the physical observable 
to the operator product expansion (OPE) of a T-product of currents at 
small distances. The OPE does not allow us to calculate power corrections, 
but one learns from the OPE the particular suppression
of higher twist effects -- $1/Q^4$ for the total   
$e^+e^-$ annihilation cross section
versus $1/Q^2$ for the DIS structure 
functions -- and the process-independence (universality) of the uncalculable 
higher twist matrix elements. The structure of power-suppressed effects 
in ``geniune''
Minkowskian quantities is much less understood. There are phenomenological
and theoretical indications that in certain situations such corrections
are large -- of order $1/Q$ -- and numerically important at all energies 
available today. In this talk we
summarize the results of Ref.~1 on the structure of power
corrections and renormalons in Drell-Yan (lepton pair) production, 
and speculate whether this example teaches us a lesson of general
validity.

Drell-Yan production, apart from its phenomenological significance, is 
theoretically interesting because it
is the simplest hard process with two large, but disparate scales, 
if we consider the production of the Drell-Yan pair (or heavy 
vector boson) close to the kinematical threshold $z\sim 1$ where $z=Q^2/s$, 
$Q^2$ being the mass of the pair and $s$ the total cms energy of 
the colliding partons.
In this situation gluon emission into the final state
is suppressed by small phase space $(1-z)Q \ll Q$,
and causes large perturbative corrections enhanced by
Sudakov-type logarithms $\ln(1-z)$. Taking moments 
$\sigma(N,Q^2) = \int dz z^{N-1} d\sigma/dz$ and subtracting
collinear divergences by forming the ratio of the 
Drell-Yan cross section and the quark distribution squared, 
one is left with the perturbative expansion for the
logarithm of the ``hard'' cross section
\be
\ln \omega_{DY}(N,Q^2) = \omega_1\alpha_s\ln^2 N+
                     \omega_2\alpha_s^2\ln^3 N+\ldots
                     \omega_k\alpha_s^k\ln^{k+1} N+\ldots,
\label{seriesLLA}
\ee
where $\alpha_s=\alpha_s(Q)$ and we have shown terms with the leading power
of the large logarithm at each order of $\alpha_s$. 
In leading logarithmic approximation (LLA), these terms are resummed to 
all orders by the elegant formula
\be
 \ln \omega_{DY}(N,Q^2) = \frac{2C_F}{\pi}
 \int_0^1 dz\frac{z^N-1}{1-z}\int_{Q^2(1-z)}^{Q^2(1-z)^2}\!\!
  \frac{dk^2_\perp}{k^2_\perp}\,\alpha_s(k^2_\perp).
\label{LLA}
\ee           
Expansion of the running coupling under the integral in powers of 
$\alpha_s(Q^2)$ generates a perturbative series which correctly reproduces
all coefficients in Eq.~\ref{seriesLLA}. Eq.~\ref{LLA} 
takes into account soft and
collinear gluon emission, and can be improved systematically by
including next-to-leading logarithms etc. The leading 
soft-collinear  contribution 
has a geometrical interpretation in terms of the cusp (eikonal)
anomalous dimension; it is  universal and appears in  
resummation of threshold corrections to any hard QCD process. 
  
However, the resummed cross section in the LLA now 
appears to be sensitive to the infrared (IR) region at the level 
of $1/Q$ corrections \cite{CON94a,KOR95}. Indeed, remove gluons with 
energy $Q(1-z)/2$ less than $\mu \sim \Lambda_{QCD}$ by 
inserting the appropriate 
$\theta$-function in the $z$-integral. A simple calculation shows that 
the cross section changes then by $\sim \mu N/Q$. Given 
this sensitivity to the IR region, one would 
suspect that geniune power corrections of this magnitude exist. 
The same conclusion can be obtained 
from considering divergences of the perturbative expansion of Eq.~\ref{LLA} 
in large orders (renormalons). One easily checks, however, that the dangerous 
IR contributions correspond to terms with less logarithms of $N$  
than are resummed by the LLA and thus are beyond the accuracy 
to which Eq.~\ref{LLA} has been derived.
Thus this evidence is by itself not conclusive.  
A more accurate analysis will clarify
two questions:

\begin{itemize}
\item{} Does the IR sensitivity (of order $1/Q$) 
  of the LLA resummed cross section 
  represent the `true' magnitude of power corrections to the Drell-Yan 
  process or is it artificially introduced by resummation, that is 
  by the procedure that separates those regions of higher order 
  Feynman diagrams which give rise to large logarithms?
\item{} If the exact Drell-Yan cross section has no $1/Q$ IR contributions, 
  can the resummation of large perturbative  corrections to all orders 
  be made consistent with the IR behaviour of finite-order Feynman diagrams? 
  Or is the (spurious) $1/Q$ sensitivity found in LLA intrinsic 
  and unavoidable for resummation of threshold corrections?
\end{itemize}

\section{IR sensitivity of LLA and soft gluon emission at large angles} 

We provide evidence \cite{BBDY,DMW,AZ} that the Drell-Yan cross section
is free from $1/Q$ IR contributions.
To one-loop accuracy this statement can be checked by an explicit
calculation with a small gluon mass $\lambda$ as regulator.
Nonanalytic terms in the small-$\lambda$
expansion correspond to higher twist contributions \cite{BBZ,BBB}. 
A textbook calculation gives \cite{BBDY}
at $N\gg 1$
\be
  \omega_{DY}(N,Q^2,\lambda^2)-\omega_{DY}(N,Q^2,0) =
  \frac{C_F \alpha_s}{\pi}\frac{N^2\lambda^2}{Q^2}\ln(\lambda^2/Q^2)+
  O\!\left(\frac{1}{N},\frac{N\lambda}{Q} \right).
\ee
Note that the suspected linear term $\sqrt{\lambda^2/Q^2}$ is absent; the
leading IR contribution is of order $N^2\lambda^2/Q^2$. The $N^2$ 
enhancement signals that the power corrections are determined by the
{\em smaller} of the two large scales: $Q/N$ is the 
moment space analogue of the energy available for gluon emission. 
In terms of the energy fraction $z$ the relevant scale is $(1-z)Q$ rather 
the mass of the Drell-Yan pair $Q$.

To understand the apparent $1/Q$ sensitivity of the LLA, it is 
instructive to consider the structure of the
one-loop integral for soft gluon emission
\be
 \omega_{DY} \sim \int\frac{d^3k}{2k_0}\delta[(p_1+p_2-k)^2-Q^2]
\left|{\cal M_{DY}}\right|^2. 
\ee
The matrix element (in LLA) is $\left|{\cal M_{DY}}\right|^2 
\sim 2Q^2/k_\perp^2$
and it is convenient to rewrite the phase-space integral as
$$
\int \frac{d^3k}{2k_0} \sim \int\frac{dk_\perp^2}{\sqrt{k_0^2-k_\perp^2}}\,.
$$
We now take a massless gluon, and to avoid collinear divergences
introduce an explicit cutoff on the minimal transverse momentum 
$k_\perp>\mu$. Remembering that $k_0=\sqrt{s}(1-z)/2$ and taking moments
we get 
\be\label{DYphasespace}
 \omega_{DY}^{soft} = \frac{2C_F}{\pi}\int_0^{1-2\mu/Q} dz\, z^{N-1}
\int_{\mu^2}^{Q^2(1-z)^2/4}\frac{d k^2_\perp}{k^2_\perp}\frac{1}
{\sqrt{(1-z)^2-4k^2_\perp/Q^2}}
\ee
The crucial point is now that the LLA corresponds to resummation of
soft and {\em collinear} emission, that is the leading logarithms 
(and in fact the next-to-leading as well) come from the integration 
region of small gluon transverse momentum compared to its energy
$k_\perp \ll k_0 = Q(1-z)/2\ll Q$. Thus, for summation of
large logarithms, it is safe to neglect the term $4 k_\perp^2/Q^2$
under the square root, so that 
\be
 \omega_{DY}^{soft+coll.} = \frac{2C_F}{\pi}\int_0^{1-2\mu/Q} dz\,
\frac{ z^{N-1}-1}{1-z}
\int_{\mu^2}^{Q^2(1-z)^2/4}\frac{d k^2_\perp}{k^2_\perp},
\ee
where we have replaced $z^{N-1}$ by $z^{N-1}-1$  to
take into account virtual gluon exchange. 
Taking the integrals, we get the expected double logarithm but 
also the linear term in $\mu/Q$ discussed in Sect.~1.

However, this IR contribution of order $1/Q$  comes from the
end-point integration region $1-z\sim \mu/Q$ (where $k_\perp\sim k_0$) 
in which neglect of
the $k_\perp^2/Q^2$ term under the square root is not justified.
In fact, the square root cannot even be expanded in $k_\perp^2/
(Q^2 (1-z)^2)$
since this would generate increasingly singular contributions. Instead, the
integral must be taken exactly. When this is done \cite{BBDY}, all
$\mu/Q$ terms disappear.

The physical reason for the enhanced IR sensitivity of the resummed
cross section in LLA is that we neglected soft gluon radiation
at large angles $k_\perp\sim k_0$. To recover the correct
IR behavior, the phase space integral for soft gluon 
emission has to be taken exactly; the common collinear approximation is
sufficient for summing logarithms to leading and next-to-leading 
logarithmic accuracy
but is misleading for the analysis of power-suppressed effects.

\section{Resummation of soft emission and Wilson lines}

Exponentiation of large logarithms occurs for both, 
collinear emission and soft gluon emission, including emission 
at large angles. Thus, after a complete treatment of  
resummation of soft gluon emission,     
the apparent $1/Q$ sensitivity in LLA should be compensated by 
other conributions to the exponent. The best-known generalization
of the LLA formula Eq.~\ref{LLA} was given by Sterman \cite{STE87}
\bea\label{form11}
\ln \omega_{DY}(N,\alpha_s) &=& \int_0^1 d z\,\frac{z^{N-1}-1}
{1-z} \Bigg\{2\int_{Q^2 (1-z)}^{Q^2 (1-z)^2} \frac{d k_\perp^2}{k_\perp^2}\,
\Gamma_{\rm cusp}(\alpha_s(k_\perp)) \nonumber\\
&& + B(\alpha_s(\sqrt{1-z} Q))
 +\,C(\alpha_s((1-z) Q)\Bigg\} + {\cal O}(1)
\eea 
This expression involves three ``anomalous dimensions''
 $\Gamma_{\rm cusp},B$ and $C$.
The LLA corresponds to taking into account the leading term $O(\alpha_s)$ 
in the expansion of $\Gamma_{\rm cusp}$ and neglecting $B,C$. The 
next-to-leading logarithmic accuracy requires two terms in 
$\Gamma_{\rm cusp}$ and the leading $O(\alpha_s)$ terms in $B,C$. 
In general, higher order corrections to $\Gamma_{\rm cusp},B,C$ generate
contributions with less and less powers of $\ln N$ for a given
power of $\alpha_s$.

The three terms in Eq.~\ref{form11} have the following origin:
The function ``$B$'' comes from subtracting the DIS cross section 
(squared) and is irrelevant for our discussion.
The term with a double integral resums soft
and collinear gluon radiation to all orders. This contribution
is universal for all hard processes.
Finally, ``$C$'' corrects for soft gluon radiation at large angles and 
is process-dependent. 
This term starts with $O(\alpha_s^2)$ 
in agreement with the common wisdom that the radiation at large angles 
is suppressed by two powers of $\ln N$, but taking it into account 
can be crucial to recover the correct IR behavior.

Thus, the $1/Q$ IR sensitivity of the generalized LLA 
expression given by the first term in Eq.~\ref{form11} should be  cancelled
by the ``$C$''-term. To test how this happens, one needs some 
approximation to calculate the anomalous dimensions to all orders
in perturbation theory. A convenient formal parameter is $N_f$, the number
of light fermion flavors. The leading contribution in the
large-$N_f$ limit corresponds to a chain of fermion loops inserted into
the single gluon line. Taking into account the chain of loops has two
effects: First, it generates the correct argument of the running coupling
in Eq.~\ref{form11} which is the gluon transverse momentum. [As usual, 
we tacitly assume that $N_f$ can be used to reconstruct the 
full one-loop running, determined by $\beta_0$.] Second,
counter\-terms for individual fermion loops produce non-trivial
anomalous dimensions  $\Gamma_{\rm cusp},B,C$ to all orders in $\alpha_s$.
The corresponding calculation is technical and can be found in Ref.~1.
The result is that all $1/Q$ IR contributions generated by the 
generalized LLA term are cancelled by the IR contributions related to
factorial divergence of the perturbation expansion of ``$C$''.
This tells us that although the resummation formula Eq.~\ref{form11} is 
valid, it involves significant cancellations between different
contributions and the true IR behaviour is restored only after summation of
an {\em infinite} number of terms in the expansion of ``$C$''. 
This contradicts the logic of resummation to resum an 
infinite number of large logarithms by calculating 
only a finite number of terms in the anomalous dimensions.

The simplest remedy would be to use the exact phase space factor 
for the one-loop gluon emission and replace the first term in 
Eq.~\ref{form11} by 
\be\label{form11mod}
\ln \omega_{DY}(N,\alpha_s) = 2\!\int_0^1 \!\!d z\,[z^{N-1}-1]
\int_{Q^2 (1-z)}^{Q^2 (1-z)^2}\! \frac{d k_\perp^2}{k_\perp^2}\,
\frac{\Gamma_{\rm cusp}(\alpha_s(k_\perp))}{\sqrt{(1-z)^2+4k_\perp^2/Q^2}}
+\ldots
\ee  
With this substitution the $1/Q$ IR sensitivity disappears and the 
functions $B,C$ are  modified 
starting $O(\alpha_s^2)$ so that the undesired 
behavior of $C$ in large orders is removed.

A different approach to soft gluon resummation emphasizes the 
renormalization of Wilson lines \cite{KM}. Its theoretical advantage 
is the operator language that avoids the separation of 
small-angle and large-angle soft emission. $1/Q$ IR contributions 
never appear. The starting point is the well-known fact that 
soft gluon emission from a fast quark can be described by a
Wilson line operator along the classical trajectory of the quark. 
The product of Wilson lines for the annihilating 
quark and antiquark is denoted by $U_{\rm DY}(x)$, 
where $x$ is the annihilation space-time 
point. Up to corrections which vanish 
as $z\to 1$, the partonic Drell-Yan cross section is given by \cite{KM}
\be\label{softfactorization}
\omega_{\rm DY}(z,\alpha_s) =  H_{\rm DY}(\alpha_s)\, 
W_{\rm DY}(z,\alpha_s)\,.
\ee
$H_{\rm DY}=1+{\cal O}(\alpha_s)$ is a short-distance dominated function,
independent of $z$.
$W_{\rm DY}$ is the square of the
matrix element $\langle n|T U_{\rm DY}(0)|0\rangle$, 
summed over all final states:
\be
W_{\rm DY}(z,\alpha_s) = \frac{Q}{2}\int_{-\infty}^
\infty\frac{d y_0}{2\pi}\,e^{i y_0 Q (1-z)/2}\,
 \langle 0|\bar{T}\,U^\dagger_{\rm DY}(y)
\,T\,U_{\rm DY}(0)|0\rangle
\ee
The Fourier transform is taken with respect to the energy of soft
partons and $y=(y_0,\vec{0})$.

The crucial observation is that the Wilson line depends only on
the ratio $(\mu N)/(Q N_0)$ (taking moments of $W_{\rm DY}(z,\alpha_s)$),
where $\mu$ is a
cutoff separating soft and hard emission (the renormalization scale
for the Wilson line) and $N_0$ is a suitable constant.
Hence the $N$-dependence of the Drell-Yan cross section in the soft 
limit can be obtained from the $\mu$-dependence of Wilson lines, 
which is given by the renormalization group equation (here 
$\alpha_s=\alpha_s(\mu)$)
\be\label{evolutionequation}
{}\!\!\left(\!\mu^2\!\frac{\partial}{\partial\mu^2} + \beta(\alpha_s)
\frac{\partial}{\partial \alpha_s}\!\!\right)\,\!\ln W_{\rm DY}\!
\!\left(\frac{\mu^2 N^2}{Q^2 N_0^2},\alpha_s\!\right) =
\Gamma_{\rm cusp}(\alpha_s)\,\ln\frac{\mu^2 N^2}{Q^2 N_0^2}
+ \Gamma_{\rm DY}(\alpha_s)\,.
\ee
It involves two anomalous dimensions $\Gamma_{\rm cusp}(\alpha_s)$ 
and $\Gamma_{\rm DY}(\alpha_s)$ related to the cusp 
and to presence of light-like segments on the Wilson line, respectively. 
The general solution of Eq.~\ref{evolutionequation} is given by
\bea\label{sol1}
\lefteqn{\ln W_{\rm DY}\!
\left(\frac{ N^2}{ N_0^2},\alpha_s(Q)\right)=
\ln W_{\rm DY}(1,\alpha_s(Q N_0/N))+} 
\nonumber\\&&
+\!\int_{Q^2 N_0^2/N^2}^{Q^2}
\!\frac{d k_\perp^2}{
k_\perp^2}\left[\Gamma_{\rm cusp}(\alpha_s(k_\perp)\,\ln\frac{k_\perp^2 N^2}{
Q^2 N_0^2} + \Gamma_{\rm DY}(\alpha_s(k_\perp))\right],
\eea
where we set $\mu=Q$.
The inhomogeneous second term in Eq.~\ref{sol1} can be rewritten (identically)
in a more familiar form, which resembles the first term in Eq.~\ref{form11}:
\be
 2\int_0^{1-N_0/N}\!\!\frac{d z}{1-z}\!\Bigg[\int_{(1-z)^2 Q^2}^{Q^2}
\frac{d k_\perp^2}{k_\perp^2}\,\Gamma_{\rm cusp}(\alpha_s(k_\perp))
+\,\Gamma_{\rm DY}(\alpha_s((1-z) Q)\Bigg]
\ee
Note presence of the 
{\em initial condition} $W_{\rm DY}(1,\alpha_s(Q N_0/N))$. 
Its expansion in $\alpha_s$
produces subdominant logarithms $\alpha_s^k\ln^{k-1} N$ which can be absorbed
into a redefinition of $\Gamma_{\rm DY}$:
\be\label{redef}
\Gamma_{\rm DY}(\alpha_s) \longrightarrow \tilde{C}(\alpha_s)=
\Gamma_{\rm DY}(\alpha_s)
- \beta(\alpha_s)\frac{d}{d\alpha_s}\,\ln W_{\rm DY}(1,\alpha_s).
\ee
$\tilde{C}(\alpha_s)$ starts at order $\alpha_s^2$.
It does not affect resummation of large
logarithms in $N$ to next-to-leading accuracy $\alpha_s^k\ln^k N$.

It remains to subtract the DIS cross section, which can also be 
implemented in the language of  Wilson lines, see Ref.~9 for details.
Finally, we get
\bea\label{form2}
\ln \omega_{\rm DY}(N,\alpha_s) &=& -\int_0^{1-N_0/N}\!\!
d z\,\frac{1}{1-z} \Bigg\{2\int_{Q^2 (1-z)}^{Q^2 (1-z)^2}
\frac{d k_\perp^2}{k_\perp^2}\,\Gamma_{\rm cusp}(\alpha_s(k_\perp))
\nonumber\\&&{} +
\tilde{B}(\alpha_s(\sqrt{1-z} Q))
 +\,\tilde{C}(\alpha_s((1-z) Q))\Bigg\} + {\cal O}(1)\,.
\eea 
($N_0=\exp(-\gamma_E)$ in the $\overline{\rm MS}$ scheme.)
This form of the resummed cross section is as 
legitimate in the framework of the perturbation theory as 
the more conventional expression in Eq.~\ref{form11}. They have different 
properties, however, as far as sensitivity to the IR behavior 
of the coupling is concerned, which becomes  
important when the anomalous dimensions $\Gamma_{\rm cusp},\ldots$ are
truncated to finite order.
Since the region of very large $z\to 1$ 
is removed in Eq.~\ref{form2}, this expression shows no IR sensitivity 
at all unless $N>Q/\Lambda_{\rm QCD}$. 
Loosely speaking, this is so because the Wilson line approach
treats small- and large-angle gluon emission in a coherent way. 
Because this technique can also treat subleading logarithms in a 
systematic way, it is preferred over, for example, the modification 
of the phase space as in Eq.~\ref{form11mod}. 

It is natural to expect (and explicit calculation \cite{BBDY} in
the large-$N_f$ limit confirms this) that the anomalous dimensions  
$\Gamma_{\rm cusp}(\alpha_s)$ 
and $\Gamma_{\rm DY}(\alpha_s)$ in the $\overline{MS}$ scheme 
are analytic functions of the coupling
at $\alpha_s=0$. Then, all power 
corrections to the resummed cross section 
(to all orders in $N\Lambda_{\rm QCD}/Q$) 
originate exclusively from the initial condition
for the evolution equation for the Wilson line, and are not
created (or modified) by the evolution, i.e. by soft gluon 
resummation. 
Thus, if the resummation of soft gluon emission is done coherently
for all angles, the only effect of soft gluons on power 
corrections is a {\em change of scale}, the replacement 
$Q\to Q/N$ as the parameter of the power expansion.
This suggests that, in general, there is no reason to suspect
new nonperturbative contributions in resummed cross sections
as compared to finite-order calculations. The conclusion that 
power corrections to Drell-Yan production are suppressed by $1/Q^2$ 
is then consistent with the analysis of power corrections 
at tree level by Qiu and Sterman \cite{QIU91}. 

The redefinition in Eq.~\ref{redef} transforms the 
IR sensitivity (and potential power corrections) of the 
initial condition for the evolution equation for
the Wilson line into IR sensitivity  of the function
$\tilde{C}$ in Eq.~\ref{form2}. As mentioned above, this function 
becomes important precisely when one starts to be sensitive to 
gluon radiation at large angles, and the conclusions on power corrections 
depend sensitively on this region. Because of this, 
we are sceptical that universality of nonperturbative corrections to 
resummed cross sections could be deduced from the universality of 
soft-collinear gluon emission as embodied by the eikonal (cusp) 
anomalous dimension, an idea originally put forward in Refs.~3,11.
In the Wilson line technique the solution
of the evolution equation never involves the QCD coupling integrated 
over the IR Landau pole as long as $N<Q/\Lambda_{QCD}$. Since this inequality 
sets the boundary for a perturbative treatment anyway, Eq.~\ref{form2} 
(which coincides with the resummation procedure used in Ref.~12 to 
next-to-leading logarithmic accuracy) is suited for all moments 
that can be treated in a power expansion in $1/Q$. 
   
\section{Top quark production at the TEVATRON}

In the light of our discussion, let us consider recent results for the
resummed top quark production 
cross section, which we summarize
in Table~1. We concentrate on the comparison of two new calculations
\cite{tt2,tt3}. Both assumed $m_t =175$ GeV and used the same parametrisations
for the structure functions. Hence the difference is entirely due to
different resummation prescriptions. The difference in central values
is of order 15\%, compared to $\sim 10\%$ renormalization 
scale dependence and $\sim 5\%$ due to uncertainty in the structure 
functions. Apparently, resummation causes the largest ambiguity. 
Note that the resummed cross section of Ref.~14 practically coincides
with the strict $O(\alpha^2)$ result. Thus, in Ref.~14 the resummation 
of $\ln N$ terms has a negligible effect,
while the resummation in Ref.~13  produces a $10-15\%$ enhancement.
Since both procedures sum all leading logarithms 
(in the sense of Eq.~\ref{LLA}),
 the difference is entirely due to terms with 
less powers of logarithms which are beyond the accuracy
of the resummation in the strict sence. Unless we 
can prefer one particular resummation procedure, the difference 
would have to be considered as the present theoretical uncertainty. 
Our discussion of Drell-Yan production suggests the criterium 
that resummation procedures should not introduce power corrections 
(factorial divergence in large orders) which are not already 
present in finite order approximations. From this point of view, 
we are led to prefer the prescription used in Ref.~14, which starts 
from Eq.~\ref{form2}. 

\begin{table}[t]
\caption{Resummed cross section for the $t\bar t$ production at the 
TEVATRON. $m_t=175\,$GeV, MRSA' parton distributions for the 
central value quoted. \label{tab:exp}}
\vspace{0.4cm}
\begin{center}
\begin{tabular}{|c|c|c|}
\hline
& &  \\
Ref. & $\sigma_{t\bar t}$, pb & Uncertainty 
\\ & &  \\ \hline
LSN \cite{tt1} & 4.95 & 4.53 -- 5.66 \\
BC \cite{tt2} & 5.32 & 4.93 -- 5.40 \\
CMNT \cite{tt3} & 4.75 & 4.25 -- 5.00 \\
\hline
\end{tabular}
\end{center}
\end{table}

The major numerical difference between Ref.~13 and Ref.~14
comes from a different procedure to implement the inverse Mellin 
transformation from moment space to momentum space. 
The subtle problem here is to which extent one can avoid contributions 
 of very large moments 
$N\ge Q/\Lambda_{QCD}$, which strictly speaking can not be 
treated by short-distance methods. 
This problem is somewhat similar 
to the problem of analytically  
continuing perturbative QCD predictions from Euclidian in Minkowski
space, relevant, for example, in connection with the $\tau$-lepton 
hadronic width. 
The particular way of performing the analytic continuation
becomes important when one uses a {\em resummed} coupling constant, 
and the guiding principle 
proves to be to avoid the region $Q<\Lambda_{QCD}$ 
in the complex $Q$-plane, where no short-distance treatment is 
possible. 
If the region $Q< \Lambda_{QCD}$ is not avoided, one may introduce 
spurious $1/Q^2$ power corrections to the decay width \cite{BBB}, 
which are absent
in the OPE. Similarly, Catani {\em et al.} find \cite{tt3} that the
inverse Mellin transform of the resummed cross section in moment space
has to be done by exact numerical integration in the complex $N$ plane,
avoiding the  region ${\rm Re}\,N\to \infty$ where the IR singularity in
the running coupling becomes important. Failure to avoid this region
may result in spurious effects of order $(\Lambda_{QCD}/Q)^{\sim 0.3}$.
  
Within the approach of Ref.~13 this problem is somewhat masked by
using a resummation formula similar to Eq.~\ref{form11}, in which 
the  sensitivity to the IR behavior of the coupling is of order $1/Q$ 
for {\em any} $N$. As explained above, this IR sensitivity is an 
artifact of truncating the anomalous dimensions to finite order. 
One also notes that applying the prescription of Ref.~13 to 
Drell-Yan production requires to introduce a phenomenological 
$1/Q$-correction, that mainly seems to cancel the $1/Q$-effects 
generated by the resummation prescription \cite{ALV94}.

The difference between various resummation 
procedures should also be perceptible in high-$p_\perp$ jet production 
at the TEVATRON, and a theoretical understanding of this difference 
might be one aspect in understanding the apparent excess of 
large-$p_\perp$ jets seen by CDF.
 
\section{$1/Q$ IR sensitivity of thrust}

The Drell-Yan cross section appears to have no $1/Q$ power 
corrections. This is not generally the case for any quantity. 
There are good reasons to suspect the existence of $1/Q$
nonperturbative hadronization corrections to event shapes 
observables in $e^+e^-$ annihilation \cite{WEB94}. Unlike the 
Drell-Yan cross section, these observables
cannot be expressed directly in terms of Feynman diagrams, 
and are obtained by
integrating the QCD amplitudes with certain weight functions such as to 
emphasize a particular final state configuration. 
These weight functions generally destroy the 
balance of gluon emission at small and large angles,
and make these observables sensitive to nonperturbative momentum
flow at large angles. As a consequence $1/Q$ corrections are 
invariably expected for event shapes. 
For example, the average thrust $\langle 1-T \rangle$
of the final state
is computed to leading order in 
$\alpha_s$ by inserting the factor
\be
  1-T = (k_0-\sqrt{k_0^2-k_\perp^2})/Q
\ee
into the phase-space integral for gluon emission, which has a 
structure identical to the Drell-Yan cross section 
in moment space, see (\ref{DYphasespace}).
The above factor suppresses small-angle emission but causes 
$1/Q$ IR sensititvity.

An interesting speculation is whether  
$1/Q$ corrections to event shapes are universal in the sense that
they can  be related to a single nonperturbative parameter 
\cite{DOK95,AKH,KS2}. Although, due to importance of large angle 
emission, this parameter would not be related to the universal 
cusp anomalous dimension, the hypothesis makes sense  
so long  as the underlying physical process is the same 
for all event shapes. Strictly speaking, the answer seems to be 
negative, since $1/Q$ corrections also occur outside the two-jet 
region\cite{NAS95} and higher-order corrections are not suppressed, 
because $\alpha_s$ is evaluated at low scale, so that it is not counted.
One may still argue in favour 
of {\em approximate} universality \cite{DMW,Dok}, 
if the coupling stays finite and reasonably small in the infrared. 
This purely phenomenological hypothesis could in principle 
be subjected to experimental tests. This, in fact, seems very hard 
in practice, because of poor control of higher orders of the perturbative
series. One may suspect that the largest part of the hadronization correction
to event shapes estimated by  Monte Carlo event generators is in fact 
related to the perturbative parton cascade which can indeed be
universal to the extent that the event shape variable 
is dominated by the two-jet kinematics. 

To illustrate the difficulty in testing 
universality consider the average thrust 
$\langle 1-T \rangle$. The existing experimental data are well described
by \cite{WEB94} 
\be
   \langle 1-T \rangle(Q) = 0.335\,\alpha_s(Q) +1.02\, \alpha_s(Q)^2 +
   \frac{1 {GeV}}{Q}
\ee   
where the first two terms give the QCD calculation (to $O(\alpha_s^2$)
accuracy). The power correction is needed to gain agreement
with the data (over the entire range of $Q$).
The second order perturbative correction has a large coefficient,
indicating that the adopted scale $Q$ is in fact inadequate for this
process. The scale setting problem for event shapes is difficult. However,
as a natural first guess one can try to take $\alpha_s$ at a  scale
of order of the jet mass $M$, which is related to thrust in the 
two-jet limit by $M^2 = (1-T)\,Q^2/2$ [We take the scale $Q_*^2=(1-T)\,
Q^2/4$, since for fixed $T$ 
this is the upper limit on the gluon transverse momentum.].
Taking $\alpha_s(m_Z)=0.12$ and using the fixed-order QCD result
$\langle 1-T \rangle\sim 0.07$, we get $Q_*\sim 0.13 Q$. Fitting 
again a $1/Q$ correction to the same data, we get 
\be
   \langle 1-T \rangle(Q) = 0.335\,\alpha_s(Q_*) +
   0.19\, \alpha_s(Q_*)^2 +
   \frac{0.4 {GeV}}{Q}
\ee 
The second-order coefficient in the perturbative series has
become much smaller and the size of the required hadronization
correction has also decreased.
One might actually think of writing the power 
correction as $0.05 GeV/Q_*$. Viewed this way, the issue of universality 
becomes inseparable from the problem of determining the most appropriate 
scale for the process.

%
%Obtaining theoretical control over $1/Q$ corrections will be very 
%interesting. But understanding scale setting and 
%higher-order perturbative effects in these observables 
%might be even more important phenomenologically and is necessary 
%to test universality.  

\end{document}